\documentclass[pss,fleqn]{w-art}
\usepackage{times}
\usepackage{w-thm}
\usepackage[]{graphicx}
\begin{document}
\keywords{Magnetic semiconductors, electronic structure.}
\subjclass[pacs]{71.55.Gs, 71.70.-d, 71.70.Gm}

\title[ZB MnTe in the GW approximation]{Electronic structure of zinc-blende 
MnTe within the GW approximation}

\author[Fleszar]{A. Fleszar\footnote{Corresponding
     author: e-mail: {\sf fleszar@physik.uni-wuerzburg.de}, Phone: 
     +49\,931\,888\,5879,
     Fax: +49\,931\,888\,5141}} 
     \address{Institut f\"ur Theoretische Physik und Astrophysik,
     Universit\"at W\"urzburg, Am Hubland, 97074 W\"urzburg, Germany}
\author[Potthoff]{M. Potthoff}
\author[Hanke]{W. Hanke}

\begin{abstract}
Using the local spin-density approximation (LSDA) and the (non self-consistent)
GW approach, the (quasi-particle) band structure is calculated for MnTe in 
zinc-blende geometry.
Different parameters characterizing the electronic structure are computed 
for an antiferromagnetic and the ferromagnetic phase and compared with the 
experiment. 
The strong Hubbard-type repulsion on the Mn-$3d$ orbitals and the $p$-$d$ 
hybridization are seen to be responsible for substantial defects found in 
the LSDA picture. 
It is discussed to which extent these can be improved upon by means of the
GW approach.
\end{abstract}

\maketitle

\section{Introduction}

Diluted magnetic semiconductors (DMS), i.e., semiconductors
with a certain amount of diluted magnetic ions, are well-known 
materials for 30-40 years. The combination of semiconducting
and magnetic phenomena in one and the same material gives rise to 
novel magnetic, magneto-optic, and magneto-transport properties
which are interesting for technological applications as well as from 
fundamental, scientific point of view \cite{Fur}.
DMS materials have attracted a perhaps unprecedented world-wide interest 
after it was demonstrated that it might be possible to produce a ferromagnetic 
semiconductor at temperatures approaching room temperature \cite{Dietl}.
The high solubility of Mn in II-VI compounds and the possibility to prepare 
good-quality samples with almost arbitrary composition range of Mn have made
the A$^{{\rm II}}_{1-x}-$Mn$_x-$B$^{{\rm VI}}$ ternary compounds the basic 
representatives of the DMS material class. Cd$_{1-x}$Mn$_x$Te and 
Zn$_{1-x}$Mn$_x$Te are perhaps the most studied members of this group.

In a phenomenological approach to the magnetic properties of DMS materials, 
one usually describes the Mn-Mn interactions with a Heisenberg model, while 
the interaction of band electrons or holes with the localized Mn magnetic moments
is modelled by means of a Kondo Hamiltonian \cite{Fur}. The parameters
of these models, i.e., the exchange constants of the Heisenberg model,
and the $J_{sp-d}$ parameters of the Kondo Hamiltonian, have to be interpreted
as effective parameters that, via (strong-coupling) perturbation theory, can be 
traced back to the Hubbard-$U$ repulsion on the Mn-$3d$ shell and the relative 
position of the Mn-$3d$ states with respect to the valence-band maximum (VBM) 
\cite{Lar}.
In any case, phenomenological or ab initio,
the electronic and magnetic properties of DMS materials are determined in first
place by the electronic configuration of the substitutional Mn ion
and its interaction with the surrounding anions. 

Mn formally belongs to the group VII of the periodic table.
Its atomic valency is given by five $3d$ and two $4s$ electrons.
Substituted as a cation in a II-VI compound, however, Mn behaves
as a group II element. The two $4s$ electrons build bonds together
with the anions' valence $p$ electrons, while the $3d$ electrons remain
in their atomic $S=\frac{5}{2}$ high-spin configuration giving rise to
a localized magnetic moment within the semiconducting host.
Apart from its magnetic moment,
the half-filled $3d$ shell of Mn resembles the fully filled $d$ shell
of Zn, Cd, or Hg in II-VI compounds. The
interaction of the valence $sp$ electrons with the localized $d$ electrons
has an important effect on the properties of these materials. 
It considerably complicates the ab initio description of II-VI semiconductors
since an independent-particle picture becomes problematic.
At least with respect to excitation properties, this makes calculations
based on the local-density approximation (LDA) \cite{HKS} questionable.
Furthermore, the LDA is known to strongly overestimate the $sp$-$d$ 
hybridization, thus pushing up the VBM and making the fundamental energy gap 
much smaller than it would result from the "usual" neglect of many-body effects 
in LDA for standard semiconductors \cite{WZ1}.

In the case of the substitutional Mn in II-VI compounds as well as for 
pure Mn-chalcogenides, there are two factors that are relevant in
the context of $sp$-$d$ interaction and that are novel when comparing
with the normal II-VI materials. First, the energy position
of the atomic Mn-$3d$ level is at least 3 eV higher (with respect to the 
vacuum level) than the positions
of the $3d$, $4d$ and $5d$ levels of Zn, Cd and Hg, respectively.
As a result, the Mn-$d$-shell-derived bands appear in II-VI DMS 
materials or in pure Mn-chalcogenides at about 3.5-4 eV below the VBM,
i.e., they are resonant with the $sp$ bands. This is a new feature
compared with II-VI compounds where the $d$-shell-derived bands
are situated at 10-8 eV below VBM and are therefore well separated from 
the $sp$ bands. Second, the spin configuration of the Mn ions is
also relevant for the hybridization because of the Hubbard-$U$
``energy barrier'' for the spin-down (unoccupied) states. In the normal
II-VI compounds the spin-up and spin-down states are both occupied
and have the same energy position.

In this paper we discuss the band structure of zinc-blende MnTe as a
prototype material. Results obtained with the conventional local spin-density 
approximation (LSDA) \cite{HKS,LSDA} are presented and contrasted with 
many-body corrections obtained from the GW approach \cite{HL,Aul}. 
The technique used is similar to our earlier studies of the electronic 
structure of the II-VI compounds \cite{GWII-VI} and other semiconductors 
\cite{SiGW,AFGW}. 
A comparison with the case of non-magnetic II-VI materials is one of the 
goals of this paper. The band structures
of MnTe in different magnetic phases are analyzed and some
phenomenological parameters are determined. 

\section{Zinc-blende MnTe}

In nature, MnTe crystallizes in the hexagonal NiAs structure. 
It is an antiferromagnet below a N\'{e}el temperature of 307 K. 
The zinc-blende phase (ZB) of MnTe is metastable.
It is easily grown, however, with the MBE technique.
Hence, ZB MnTe bas intensively been studied for about 18 years and is oftenly 
used as a magnetic component in Mn-based diluted magnetic semiconductors. 
Apart from technological interest, ZB MnTe is also a fundamentally interesting 
material as it represents one of the few realizations of fcc antiferromagnetic
systems with dominant nearest-neighbor interactions resulting in an inherent 
magnetic frustration.
Upon cooling the system usually undergoes a structural distortion that accompanies 
the magnetic phase transition which eventually lifts the ground-state degeneracy.
This also happens with ZB MnTe which developes a tetragonal distortion of the 
order of 0.3\% below the N\'{e}el temperature \cite{Hen}.

Zinc-blende MnTe is a type-III antiferromagnet (AF3) with a N\'{e}el temperature 
of 65 K \cite{Gieb}. 
The difference to the simpler type-I structure (AF1) constists in the different
geometrical sequence of the (001) planes with the planar antiferromagnetic order: 
In the AF1 structure this is given by an A-B-A-B-A sequence of (001) planes, while the 
AF3 phase has a twice as long period: A-B-C-D-A. 
The ferromagnetic phase has higher energy and is not observed experimentally. 
Antiferromagnetic order is obviously favored by the superexchange mechanism 
\cite{Ander}, dominant in the intrinsic Mn-chalcogenides as well as in all 
II-Mn-VI DMS materials.
Another important experimental fact is the negative, i.e., antiferromagnetic
coupling of the spins of (itinerant) holes with the localized spins of Mn \cite{Fur}. 
This $J_{pd}$ coupling is very large in magnitude and the source of many characteristic 
properties of the DMS materials.

Cubic MnTe has been investigated experimentally with various techniques
and especially its magnetic \cite{Gieb,Hen} and elastic \cite{Dje}
properties are well characterized. 
In the present context, it is important to mention that the photoemission 
experiments place the occupied Mn-$3d$ level at about 
3.5 eV below the VBM \cite{PES1,PES2} whereas the experimental exchange 
splitting of Mn-$3d$ levels is about 6.9 eV \cite{IPES}. 
Theoretically, ZB MnTe was first studied
within the LSDA approximation in two seminal papers by
Wei and Zunger \cite{WZ2} and Larson et al \cite{Lar}. 
Heisenberg exchange constants for the three Mn-chalcogenides 
(MnS, MnSe and MnTe) have been derived from total-energy LSDA calculations 
(and empirically modified LSDA) by Wei and Zunger \cite{WZ3}.
In addition, structural properties of MnTe have been studied within LDA 
and the generalized-gradient approximation \cite{Gon}.

\section{Computational details}

The first part of calculations presented here has been done within the LSDA
approximation and the Kohn-Sham approach of density-functional theory \cite{HKS}.
We use the Perdew-Zunger parameterization of the LDA functional \cite{PZ}
together with the interpolation formula of von Barth and Hedin \cite{LSDA} for
intermediate spin polarization. 
Within the pseudopotential approach, a Mn$^{15+}$ pseudopotential of the Hamann 
type \cite{Ham} has been generated. Thus the whole $n=3$ atomic shell of Mn is 
taken into account in the self-consistent calculation,
i.e., including the 3s and 3p core states. For the tellurium atom
a Te$^{6+}$ pseudopotential of the BHS-type has been used \cite{BHS}.
The explicit inclusion of the Mn 3s and 3p states in calculations for solids 
is necessary for the subsequent GW calculation because of the large bare exchange
interaction within the whole $n=3$ shell \cite{RKP}. On the level
of the LSDA approximation, however, using a Mn$^{7+}$ pseudopotential
results in basically the same band structure as the Mn$^{15+}$
pseudopotential considered here -- provided that the charge density of 
inner core shells is included in form of a partial core-charge correction 
\cite{LFC}.

The Kohn-Sham equations have been solved using the mixed-basis method.
In addition to plane waves, a number of
localized Gaussians multiplied by the $s$, $p$ and $d$ spherical harmonics
are placed in the positions of the Mn atoms. Using this mixed basis
has two advantages which are particularly important within the context 
of GW calculations:
(i) The basis is universal enough to describe both, the strongly localized
core states as well as the highly delocalized excited states which both are
needed in the course of a GW calculation. 
(ii) As compared to a pure plane-wave scheme, it allows to considerably
reduce the number of plane waves.

The GW method has been applied in a similar way as in our recent calculation
for II-VI compounds \cite{GWII-VI}. The dynamical dielectric matrices
are determined on the (4,4,4) Monkhorst and Pack mesh of k-points
\cite{MP}, no plasmon-pole approximation has been applied.
The dielectric matrices are calculated in Fourier space, and only
the diagonal contribution of the self-energy has been taken into account \cite{GWII-VI}.
The calculations have been done at the experimental lattice constant 
$a=6.338\;\AA$.

\section{Results of LSDA calculations}

Figure 1 shows the resulting LSDA band structure of zinc-blende MnTe
for the ferromagnetic and the AF1 antiferromagnetic phase. 
Both magnetic structures do not appear in nature. 
Nevertheless, their analysis is highly instructive for the discussion of 
the Mn-$3d$--Te-$5p$ hybridization. 
Note that the AF1 structure is very similar to the (actually realized) AF3
structure as regards the densitiy of states, for example. 
Considering the AF1 structure, however, allows for a more transparent discussion 
and facilitates the analysis below.
The band structures shown in Fig.\ 1 agree well with previous LSDA results
\cite{WZ2}. 

\begin{vchfigure}[htb]
\includegraphics[width=.9\textwidth]{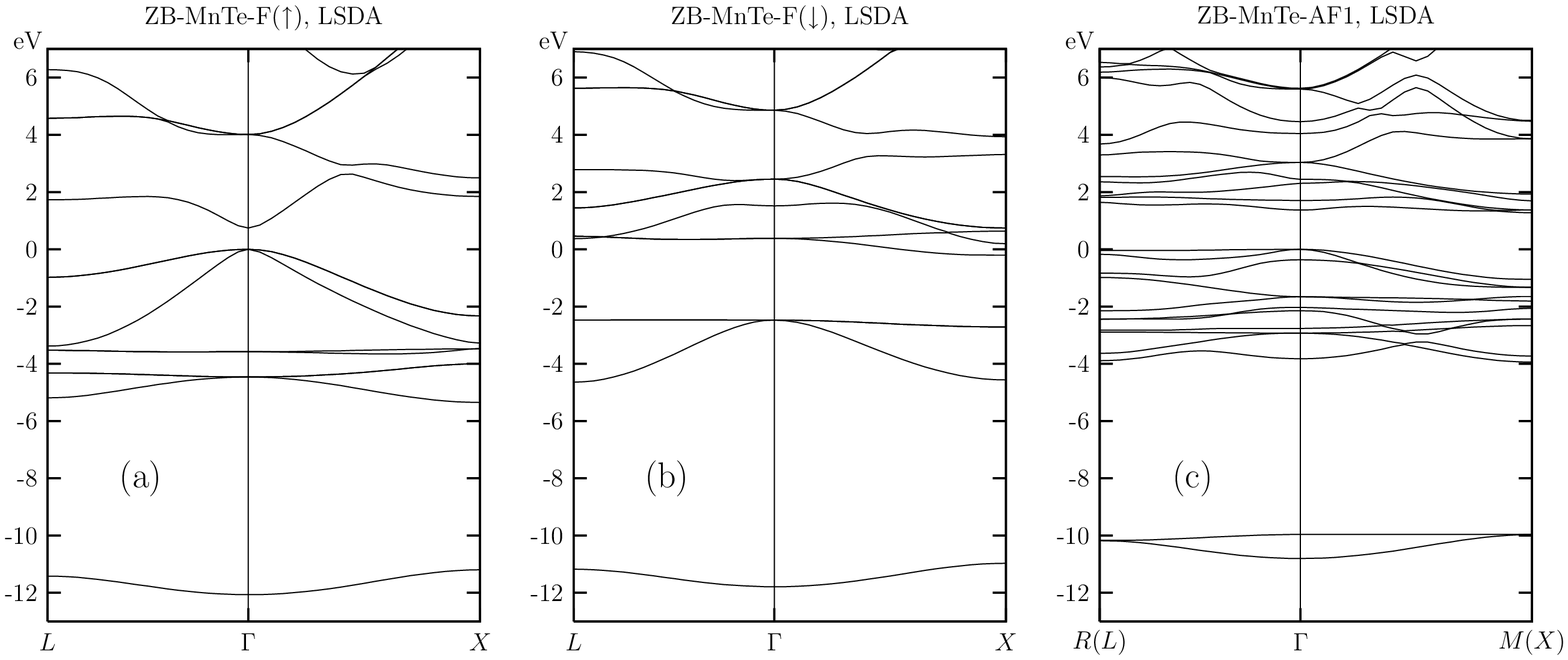}
\caption{LSDA band structure of zinc-blende MnTe in the ferromagnetic phase: 
(a) majority spin, (b) minority spin. Results for the antiferromagnetic AF1 
phase: (c).}
\label{fig:1}
\end{vchfigure}

As can be seen from Fig.\ 1, ferromagnetic ZB MnTe appears to be a metal within 
the LSDA approximation. The ``unoccupied'' minority-spin bands drop down below 
the VBM in the majority-spin channel.
By contrast, in the AF1 phase ZB MnTe turns out to be a semiconductor with 
a band direct gap at $\Gamma$ of 1.37 eV and a minimum indirect gap of 1.28 eV 
at $X$. We have also calculated the band structure for the AF3 phase.
In this case the direct band gap at $\Gamma$ is the minimum gap which 
amounts to 1.33 eV. These LSDA gaps are considerably smaller than the 
experimentally determined gap of 3.2 eV \cite{Gap1,Gap2,Gap3}. 
This should be interpreted as the usual 
(and joint) artifact of Kohn-Sham theory and LDA approximation. 

The spin-dependent density of states (DOS) for the ferromagnetic phase is presented 
in Fig.\ 2 (upper panel). The energy zero is placed at the VBM in the majority-spin 
channel. The dotted line shows the position of the Fermi energy. 
The lower panel presents the contributions to the DOS from the Mn-$3d$ orbitals. 
Projections onto states with e$_g$ and t$_{2g}$ symmetries are shown separately. 
In Fig.\ 3 the DOS for the AF1 phase is presented with a separate projection onto
two inequivalent Mn atoms in the AF1 unit cell.

There are a few features worth commenting:

(i) In the ferromagnetic phase, the VBM of the minority-spin carriers is situated
2.47 eV below the VBM in the majority-spin channel. 
This negative (and large) spin splitting of the VBM is common to all II-Mn-VI 
DMS materials and due to the negative (antiferromagnetic) $p$-$d$ exchange interaction.
This effective interaction can easily be understood as resulting from the 
Schrieffer-Wolff transformation \cite{SchWo} applied to Anderson's $p$-$d$ 
model Hamiltonian \cite{Ander1}.
The negative spin splitting of the VBM is therefore to be interpreted as a consequence 
of the strong Mn-$3d$ Hubbard-$U$ repulsion and the fact that 
the $p$ bands lie above the occupied but below the unoccupied Mn-$3d$ levels. 

The $p$-$d$ exchange is usually described by the
phenomenological parameter $N_0\beta$ \cite{Fur} defined by the relation:
$E_{\rm VBM}(\uparrow)-E_{\rm VBM}(\downarrow)=-N_0\beta x \langle S_z \rangle$,
where $x$ is the concentration of the magnetic ions ($x=1$ in our case)
and $\langle S_z \rangle$ is the average value of the total spin
on the magnetic ion. 
It is difficult to give a precise definition of the magnetic moment of a 
single atom embedded in a crystal.
For the ferromagnetic phase $\langle S_z \rangle=5/2$ in the whole unit cell 
resulting in a moment
of $5\mu_{\rm B}$ (we ignore that the magnetization actually slightly differs 
from $5\mu_{\rm B}$ within a band calculation for a metal).
The magnetic moment on the Mn-ion, however, is less than $5\mu_{\rm B}$ due to 
the hybridization. 
In our calculation, the net magnetic moment in a sphere of radius $R_{MT}=1.37 \AA$ 
(half of the distance to the next Te-neighbor) is $4.5\mu_{\rm B}$ for the ferromagnetic 
phase and $4.3\mu_{\rm B}$ for the AF1 phase. 
Thus, taking $\langle S_z \rangle=4.5/2$ one obtains an LSDA value of 
$N_0\beta=-1.1$ eV.
The experimental value for ZB MnTe is $N_0\beta^{\rm exp}=-0.88$ eV \cite{Gaj}. 
This large magnitude of $N_0\beta$ results from a strong $p$-$d$ hybridization 
between Mn-$3d$ and the anion-$p$ states. 
Within LSDA the hybridization strength is overestimated \cite{WZ1,GWII-VI}. 
This explains the larger theoretical value for $N_0\beta$ as compared to
the experiment.

\begin{figure}[htb]
\begin{minipage}[t]{.45\textwidth}
\includegraphics[width=\textwidth]{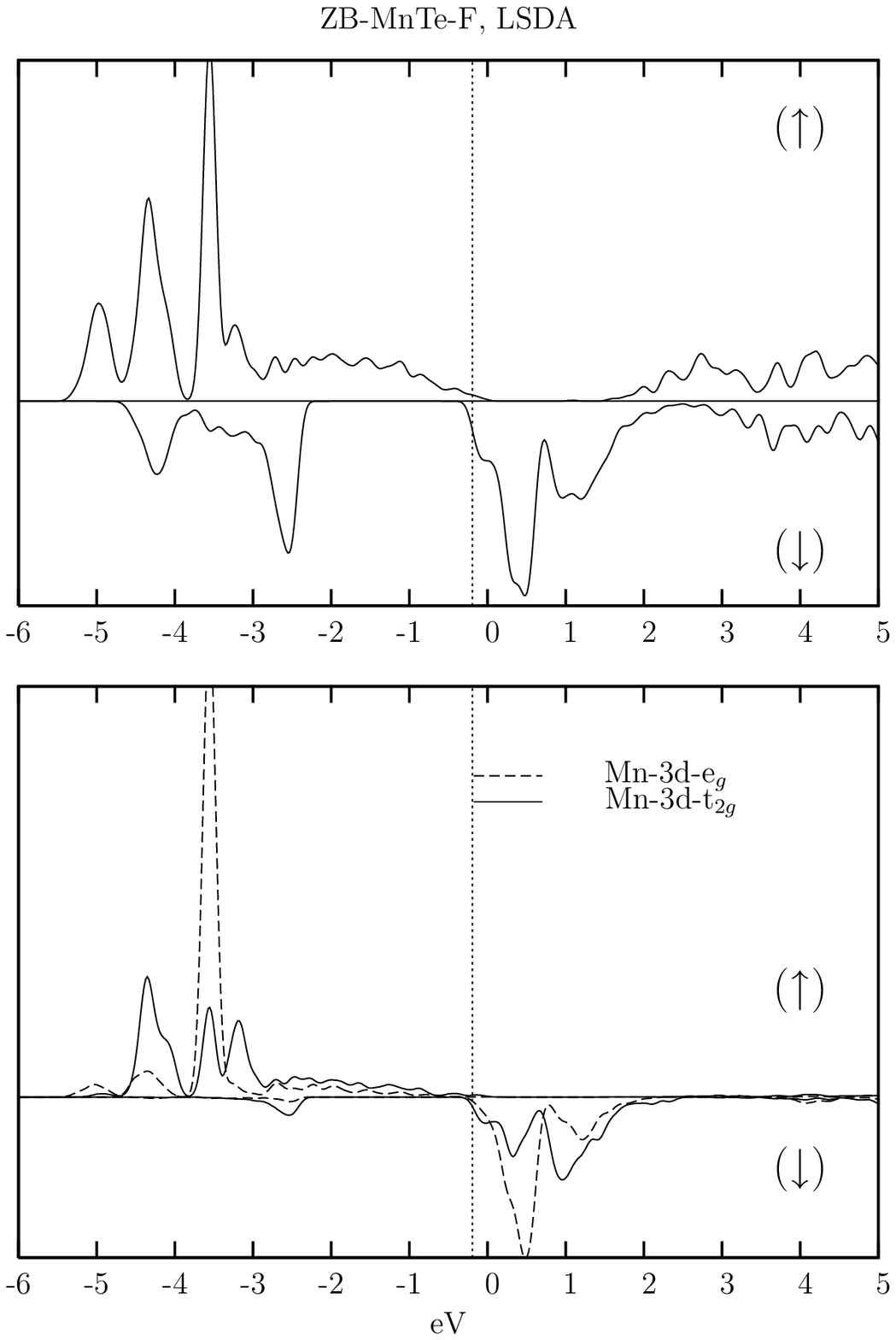}
\caption{LSDA density of states (DOS) for ZB MnTe in the ferromagnetic
phase. Upper panel: total DOS. Lower panel: contributions from
the Mn-$3d$-e$_g$ and Mn-$3d$-t$_{2g}$ states. $(\uparrow)$ 
denotes the majority-spin, $(\downarrow)$ the minority-spin channel.
Energy zero: valence-band maximum for spin ($\uparrow$). 
Dotted line: Fermi energy}
\label{fig:2}
\end{minipage}
\hfil
\begin{minipage}[t]{.45\textwidth}
\includegraphics[width=\textwidth]{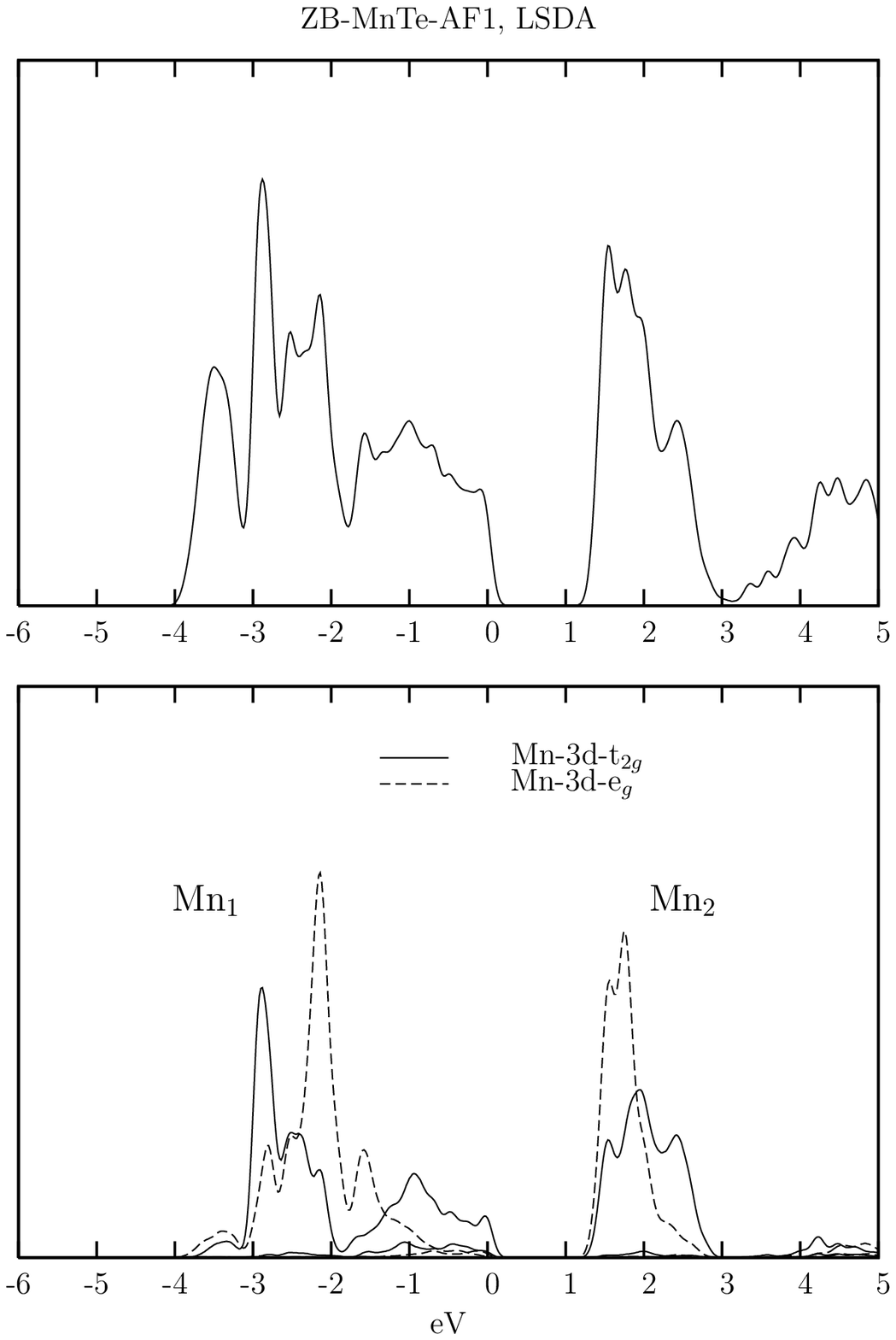}
\caption{LSDA DOS for ZB MnTe in the AF1
phase. Upper panel: total DOS. Lower panel: contributions from
the Mn-$3d$-e$_g$ and Mn-$3d$-t$_{2g}$ states. The DOS is shown for two 
inequivalent Mn atoms in the AF1 unit cell.}
\label{fig:3}
\end{minipage}
\end{figure}

(ii) The exchange splitting of the unoccupied $4s$ states at the $\Gamma$ point
is positive.
In Fig.\ 1 the conduction-band minimum (CBM) has $4s$ character for majority
spin but Mn-$3d$-e$_g$ character in the minority-spin channel. 
For minority spin, the unoccupied $4s$ states lie above Mn-$3d$-e$_g$ states 
at $\Gamma$.
Hence, the energy difference between the $s$-type bands for spin up and down is
recognized to be positive and amounts to 0.77 eV in our calculation.
This gives rise to $N_0\alpha=0.34$ eV to be compared with the experimental
value of $N_0\alpha^{\rm exp}=0.22$ eV \cite{Gaj} ($N_0\alpha$ is defined via the 
relation: $E_{4s}(\uparrow)-E_{4s}(\downarrow)=-N_0\alpha x \langle S_z \rangle$). 
A positive value of $N_0\alpha$ indicates that for the $4s$ states the spin
splitting mechanism is of a {\em direct} (or {\em potential}) type. 
Note that at $\Gamma$ the Mn-$4s$ wave function is orthogonal to the Mn-$3d$ wave 
functions and has a large amplitude at the Mn sites. 
These are conditions that favour a parallel spin configuration according to 
Hund's rule. 

(iii) Another characteristic feature is the decreasing width of the occupied 
$p$-$d$-type bands when going from the majority-spin channel in the ferromagnetic 
phase (F$(\uparrow)$), to the AF1 phase, and eventually to the minority-spin channel
in the ferromagnetic phase (F$(\downarrow)$). 
This is clearly seen in the plots of the density of states. 
The widths are 5.4 eV, 3.9 eV and 2.2 eV for the F$(\uparrow)$, AF1
and F$(\downarrow)$ case, respectively. 
The trend is caused by the decreasing number of hybridization partners for the 
Te-$5p$ states within this series. 
While for the F$(\uparrow)$ case each Te atom has four Mn-$3d$ partners of 
spin up in the local environment with tetrahedral symmetry, there are only 
two such partners for the AF1 case. 
On the other two Mn atoms (with spin-down magnetic moment) the spin-up
$3d$ orbitals, which could hybridize, are separated in energy by the 
Hubbard-$U$, and therefore hybridization is largely suppressed.
In the case of F$(\downarrow)$, all Mn-$3d$ orbitals of the same spin
are separated by the Hubbard-$U$ energy barrier.
Consequently, this gives the narrowest band.
When extracting a quantitative value for the strength of the $p$-$d$ hybridization,
which is needed for the construction of an $sp$-$d$ model Hamiltonian, one should
be aware, however, that the LDA overestimates the hybridization strength.
The size of this ``LDA error'' is difficult to estimate.

(iv) A feature related to the band widths is the position of the Mn-$3d$ states. 
The lower panels of Figs.\ 2 and 3 show the DOS projected onto the Mn-$3d$ 
contributions and resolved according to e$_g$ and t$_{2g}$ symmetry. 
For the ferromagnetic phase, not only the crystal lattice but also the Kohn-Sham 
potential is characterized by the zinc-blende (Td) symmetry. 
In this case, the Mn-$3d$ states with e$_g$ symmetry do not hybridize with the 
Te-$5p$ states, at least in the region close to the  $\Gamma$ point. 
Therefore, one observes a rather narrow peak at -3.6 eV (see dashed line in 
the lower panel of Fig.\ 2). 
By contrast, the Mn-$3d$ states with t$_{2g}$ symmetry strongly mix with the 
Te-$5p$ states resulting in a comparatively broad partial DOS with a few peaks
and ranging up to the VBM.
For zinc-blende MnTe in the AF1 phase, the symmetry of the Kohn-Sham potential 
is lower than Td (tetragonal magnetic symmetry). 
Therefore, the contribution from the e$_g$ states to the total DOS
(dashed line in the negative-energy part of Fig.\ 3) is visibly broader
than for the F$(\uparrow)$ case and composed of a few peaks around the central 
peak at 2.1 eV below the VBM. 
Hence, the difference between -3.6 eV for the F$(\uparrow)$ case and -2.1 eV 
for the AF1 case can be explained with the different available number of 
hybridization partners with the same spin. 
One should note, however, that this energy difference, 1.5 eV, is perhaps to a 
large extent due to the considerable overestimation of the $p$-$d$ hybridization 
strength within the LDA.
One could speculate that in a theory free from the LDA errors, the position of 
the e$_g$ peak could be almost independent of the magnetic phase.

(v) The exchange splitting of the Mn-$3d$ states is not straightforward to define 
because the Mn-$3d$ states undergo crystal-field and hybridization-induced 
splittings into several bands. 
We will define two exchange splittings, separately for e$_g$ and t$_{2g}$ symmetry,
by calculating the center of mass of the symmetry-projected DOS.
Furthermore, we ignore the hybridization effects and limit the considerations to 
peaks below -2 eV for the AF1 structure and t$_{2g}$ symmetry and below -2.9 eV 
for the F$(\uparrow)$ case and the same symmetry.
With these definitions we obtain the values $\Delta_{e_g}=4.3$ eV and 
$\Delta_{t_{2g}}=4.8$ eV in the ferromagnetic phase and $\Delta_{e_g}=3.9$ eV and 
$\Delta_{t_{2g}}=4.7$ eV in the AF1 phase. 
We note that the exchange splittings for the ferromagnetic phase are larger than 
for the AF1 phase as was already noticed in \cite{WZ2} but the effect is small. 
This reflects the local nature of this splitting having its origin in the 
Hubbard-$U$ on the Mn-$3d$ shell.

\section{Results of GW calculations}

Our main interest concerns the changes in the LDA band structure that are 
introduced by the GW self-energy.  
The difference between the LDA Kohn-Sham potential and the GW self-energy 
comes from the difference between the LDA exchange-correlation potential 
and the sum of the bare exchange and the dynamically screened Coulomb potential 
of an extra particle in the GW approximation.  
In perturbative GW theory, as it is used here, the LDA wave functions remain 
unchanged.
This implies that the $p$-$d$ hybridization is unchanged, i.e., an important 
ingredient affecting the band structure for all II$^B$-VI compounds. 
We conclude that, for this situation, GW corrections carry information
about the effects of screening and the effects of the true exchange,
not properly included in the LDA theory, but do not correct the
LDA errors with respect to the hybridization. In addition, the screening as well
as the bare exchange are calculated on the basis of the LDA band structure,
i.e., they are not free of LDA errors. A remedy of these problems would be 
an application of the GW method either in a self-consistent way,
or on top of a one-particle approximation free from the most severe
LDA errors. It is nevertheless instructive to see
the effects of the bare exchange and screening and their influence
on the electronic structure in different magnetic phases.

\begin{vchfigure}[htb]
\includegraphics[width=.9\textwidth]{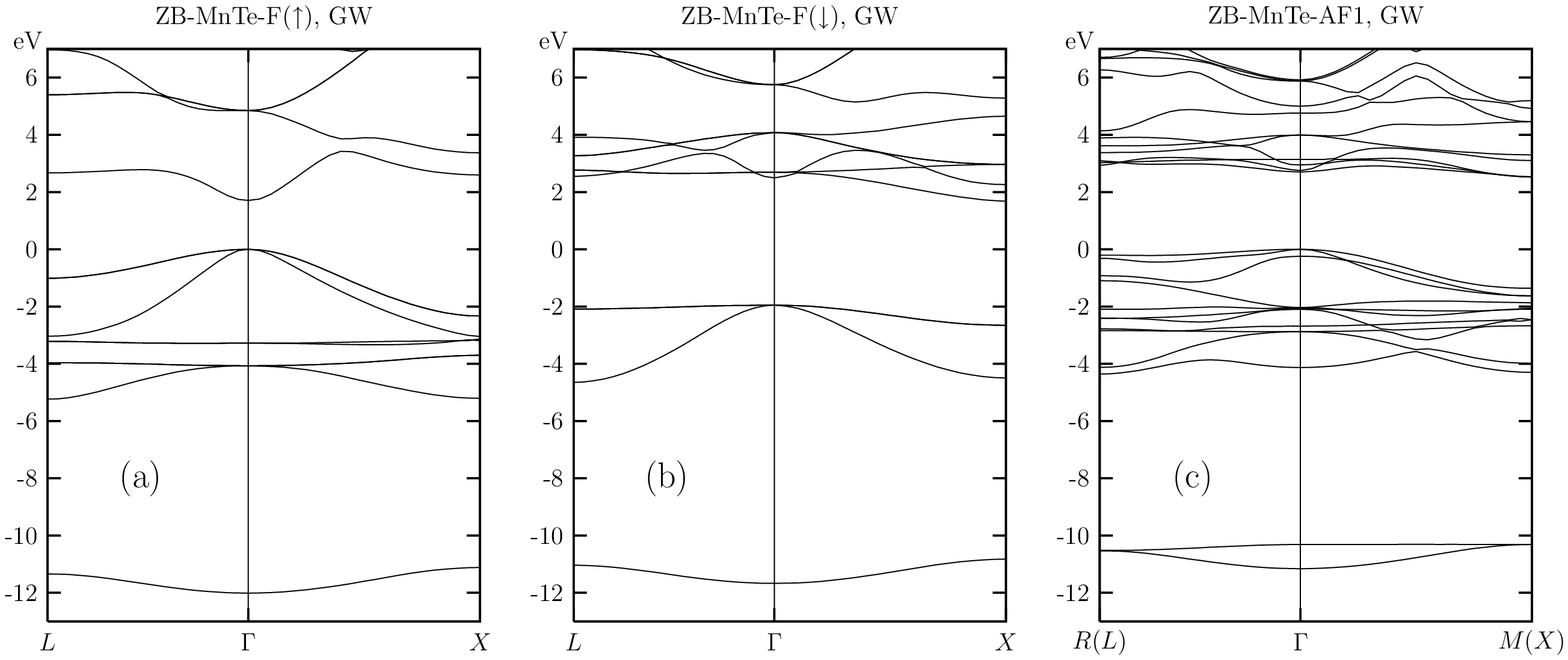}
\caption{GW band structure of zinc-blende MnTe in the ferromagnetic
phase. (a) majority spin, (b) minority spin. 
(c) Band structure for the antiferromagnetic AF1 phase.}
\label{fig:4}
\end{vchfigure}

Fig.\ 4 presents the band structure of ZB MnTe in the ferromagnetic and 
in the AF1 phase as obtained within the GW approximation. 
Figs.\ 5 and 6 show the corresponding densities of states. 
They appear somewhat smoother as compared to Figs.\ 3 and 4. 
This is due to the slightly larger energy broadening and the coarser mesh 
of k-points on which the GW density of states has been calculated.

\begin{figure}[htb]
\begin{minipage}[t]{.45\textwidth}
\includegraphics[width=\textwidth]{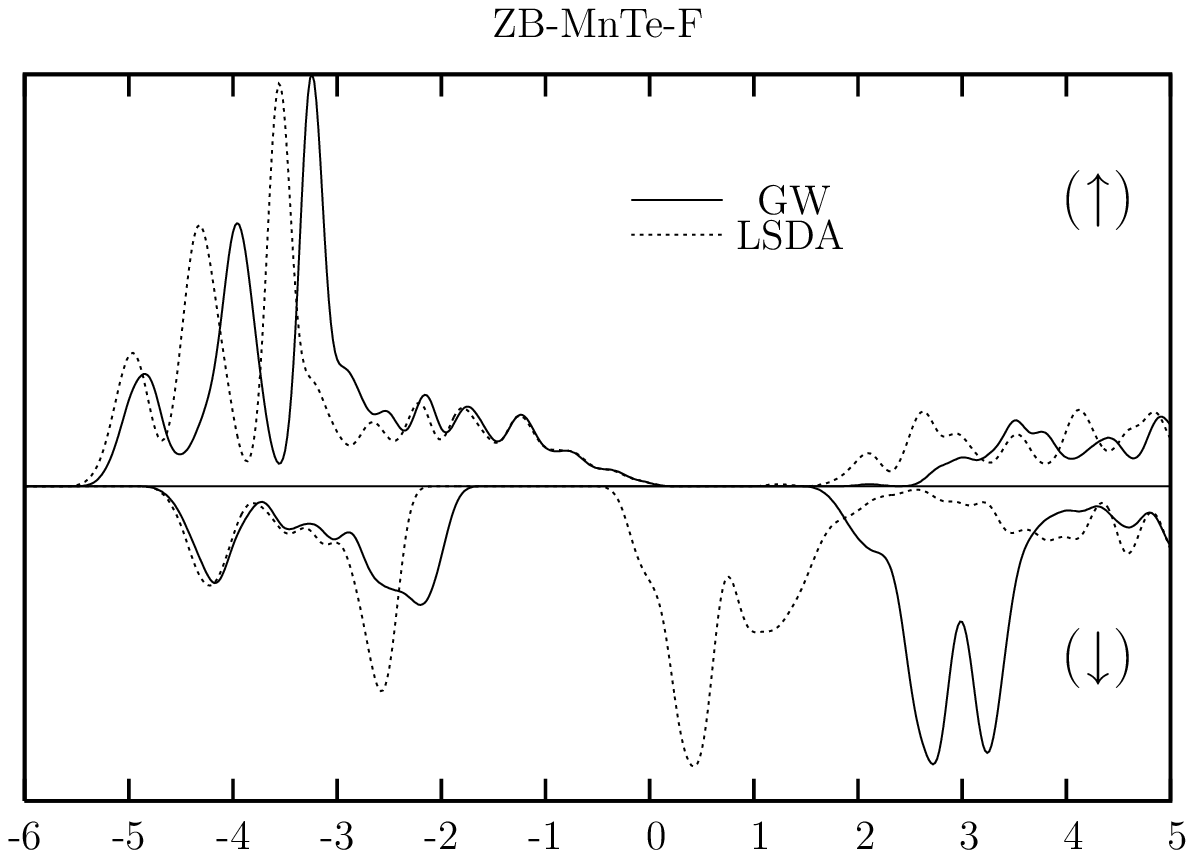}
\caption{DOS for ferromagnetic ZB MnTe.
Solid line: GW, dotted line: LSDA.}
\label{fig:5}
\end{minipage}
\hfil
\begin{minipage}[t]{.45\textwidth}
\includegraphics[width=\textwidth]{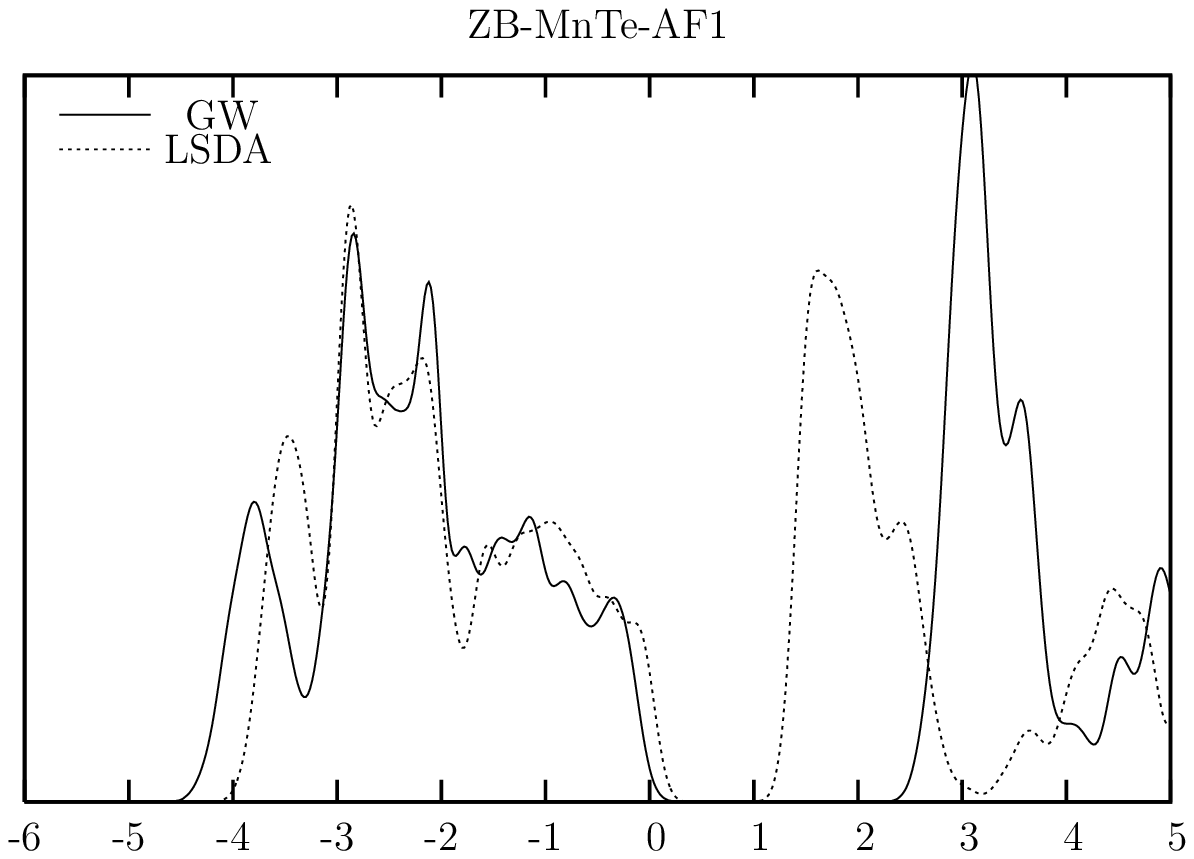}
\caption{DOS for antiferromagnetic (AF1) ZB MnTe.
Solid line: GW, dotted line: LSDA.}
\label{fig:6}
\end{minipage}
\end{figure}

The most apparent effect of the GW correction consists in an opening of the 
energy gap.
Opposed to the LSDA result, ferromagnetic ZB MnTe in no longer a metal 
but exhibits an overall positive energy gap.
The majority-spin energy gap at $\Gamma$ is 1.71 eV (0.74 eV in the LSDA). 
The gap for the minority-spin case, as measured between the 
$\Gamma_{15v}(\downarrow)$ and $\Gamma_{1c}(\downarrow)$ states, 
has increased from 4 eV in the LSDA to 4.5 eV in 
the GW approach. 
For the AF1 phase, the GW energy gap turns out to be 2.7 eV at $\Gamma$ and 
2.53 eV at $X$, to be compared with the LSDA gap of 1.37 eV at $\Gamma$ and 
1.28 eV at $X$. 
Similarly as in the LSDA, the lowest conduction band has a Mn-$3d$ character.
The $s$-type conduction band, however, has only 0.1 eV higher energy. 
Assuming that the spin-orbit (SO) splitting of the VBM is similar in MnTe 
and other tellurides, the energy gap would be $\sim 0.3$ eV smaller with SO 
effects included. 
Hence, it would be $\sim 0.8$ eV smaller than the experimental gap in the 
AF3 phase which amounts to 3.2 eV. 
This discrepancy is twice as large as for ZnTe and CdTe
where GW theory predicts an 0.4 eV smaller gap compared with the experiment
\cite{GWII-VI}. It is reasonable to ascribe this difference to the
stronger hybridization effects in MnTe as compared to II$^B$-VI compounds.

As can be seen in Figs. 5 and 6, the largest GW corrections show up for 
the unoccupied Mn-$3d$ states. 
The spin-down Mn-$3d$ states in the ferromagnetic phase are shifted
by more than 2 eV to higher energies. The unoccupied Mn-$3d$ states in the
AF1 phase are shifted by 1.5 eV. These corrections are larger than those
for diffuse $sp$ bands and can be understood as a cancellation of a 
considerable LDA error.

The occupied Mn-$3d$ states either 
retain their LDA energy position relative the VBM, as it is the case 
for the AF1 phase, or shift by $\sim 0.3$ eV {\em to higher energies} as
for the ferromagnetic phase. This is a surprising result because
one could rather expect an (opposite) downshift of the localized $d$
states which are underbound within the LDA. 
For Zn- and Cd-chalcogenides the cations' $d$ states are known to shift 
down by about 0.5-0.7 eV due to the GW correction \cite{GWII-VI}.
This shift is not sufficient to achieve agreement with the experiment,
and one of the reasons for the remaining discrepancy is the LDA
starting point of the GW procedure. We believe that this is the same
reason why the Mn-$3d$ states in the present calculation do not move 
or move upwards, respectively. 
For MnTe the $p$-$d$ hybridization is stronger than for II$^B$-VI materials. 
This stronger hybridization precludes a downshift of the Mn-$3d$ states 
in the AF1 phase and, for the ferromagnetic phase where there are even 
more hybridization partners, results in an upwards shift. 
Since GW theory tends to decrease the width of $p$-valence bands in 
II$^B$-VI materials \cite{GWII-VI}, an admixture of $p$-type
wave functions to $3d$-type states gives rise to a competition
between down and up GW shifts.

The average energy position of the Mn-$3d$ occupied states for the AF1 phase 
is the same within LSDA and GW and equal to 2.5 eV below the VBM. 
This agrees well with the results of Wei and Zunger
\cite{WZ2}. In photoemission experiments \cite{PES1,PES2} only one Mn peak
was resolved at 3.5 eV below VBM.

The GW exchange splittings of Mn-$3d$ states in the AF1 phase,
$\Delta_{e_g}=5.4$ eV and $\Delta_{t_{2g}}=6.2$ eV, should be compared
with the experimental value of 6.9 eV \cite{IPES}. For the ferromagnetic
phase, GW predicts $\Delta_{e_g}=6.2$ eV and $\Delta_{t_{2g}}=6.7$ eV,
respectively.
Since within GW the energies of quasiparticle states can be interpreted
as total-energy differences of systems with $N$ and 
$N\pm1$ particles, the calculated exchange splittings can be seen
as the GW approximation for the screened Hubbard-$U$.

The GW exchange splitting of the VBM in the ferromagnetic phase of -1.95 eV
is smaller than in the LDA. 
This results in $N_0\beta=-0.87$ eV within the GW approximation which is very 
close to the experimental value of -0.88 eV. 
For the exchange splitting of the $s$-type conduction bands at $\Gamma$,
the GW calculation yields $E_{4s}(\uparrow)-E_{4s}(\downarrow)=
-0.79$ eV. This gives rise to $N_0\alpha=0.35$ eV which, similarly as in
case of the LSDA, is larger than the experimental value $N_0\alpha^{Exp}=0.22$ eV.
One should remember, however, that the theoretical
results suffer from the uncertainty in the definition of the local
magnetic moment on the Mn ion while the experimental results have been
determined as a statistical average on systems with different Mn content.

Tables 1 and 2 collect the values of the parameters derived in this work
in comparison with previous calculations as well as with the experiment.

\begin{table}[htb]
\begin{minipage}[t]{.45\textwidth}
\caption{Energy levels and exchange splittings for ZB-MnTe in the
F phase. All values in eV.}
\label{tab:1}\renewcommand{\arraystretch}{1.5}
\begin{tabular}{lrrr} \hline
 & LDA & GW \\ \hline
$\Gamma_{15v}(\uparrow)-\Gamma_{15v}(\downarrow)$ & 2.47 & 1.95  \\

$\Gamma_{1c}(\uparrow)-\Gamma_{1c}(\downarrow)$ & -0.77 & -0.79  \\

$E_g(\uparrow): \Gamma_{1c}^{\uparrow}-\Gamma_{15v}^{\uparrow}$ & 0.74 &
1.71 \\

$E_g(\downarrow): \Gamma_{1c}^{\downarrow}-\Gamma_{15v}^{\downarrow}$ &
4.0 & 4.5 \\

$\bar{E}_{d_{e_g}}(\uparrow)$ & -3.7 & -3.4 \\

$\bar{E}_{d_{t_{2g}}}(\uparrow)$ & -3.9 & -3.6 \\

$\Delta_{e_g}$ & 4.3 & 6.2 \\

$\Delta_{t_{2g}}$ & 4.8 & 6.7 \\ \hline

\end{tabular}
\end{minipage}
\hfil
\begin{minipage}[t]{.45\textwidth}
\caption{Energy levels and exchange splittings for ZB-MnTe in the
AF1 phase. All values in eV.}
\label{tab:2}\renewcommand{\arraystretch}{1.5}
\begin{tabular}{lrrr} \hline
 & AF1-LDA & AF1-GW & Expt \\ \hline
$E_g(\Gamma)$ & 1.37 & 2.7 & 3.2$^a$  \\

$\bar{E}_{d_{e_g}}(\uparrow)$ & -2.2 &  -2.2 &  \\

$\bar{E}_{d_{t_{2g}}}(\uparrow)$ & -2.7 & -2.7 & \\

$\bar{E}_d(\uparrow)$ & -2.5 & -2.5 & -3.5$^b$ \\

$\Delta_{e_g}$ & 3.9 & 5.4 &  \\

$\Delta_{t_{2g}}$ & 4.7 & 6.2 &  \\

$U$   & 4.4 & 5.9 & 6.9$^c$ \\ \hline 
\end{tabular}
$a$-Ref.\cite{Gap1,Gap2,Gap3}, $b$-Ref.\cite{PES1,PES2}, $c$-Ref.\cite{IPES}
\end{minipage}
\end{table}

\section{Conclusions}

Our analysis of the electronic structure of zinc-blende MnTe has shown up
substantial defects of the LSDA which, at least partially, can be improved
within the GW approach:
Energy gaps are strongly improved within the GW approximation.
However, similarly to non-magnetic II-VI compounds, they are still smaller 
than in the experiment. 
A similar effect has been observed for the average exchange splitting of 
the Mn-3$d$ states (Hubbard-$U$): 
While there is a correction by 1.5 eV with respect to the LSDA result, the
splitting still turns out to be about 1 eV smaller than in the experiment. 
The binding energy of the Mn-$3d$ states remains the same within GW as 
compared to LSDA and is $\sim$ 1 eV higher than in the experiment. 
We attribute this fact to the strongly overestimated $p$-$d$ hybridization 
in the LDA theory which is not corrected within the GW approach if applied 
perturbatively. 
GW predicts the $N_0\beta$ parameter close to the corresponding experimental 
value.
On the other hand, there is no improvement with respect to $N_0\alpha$.
Concluding, the analysis shows that a non self-consistent GW approach is
inappropriate to correctly include the important $p$-$d$ hybridization effects.
In our opinion, it is mainly the LDA starting point that needs to be improved.
This would be a necessary step for a reliable description of the thermodynamics
of DMS materials derived within a model for the local electronic structure 
taken from ab initio calculations.

\begin{acknowledgement}
We would like to thank W. Szuszkiewicz for valuable discussions.
The numerical calculations have been performed at the ZAM J\"ulich.
Financial support of this work by the Deutsche Forschungsgemeinschaft 
within the Sonderforschungsbereich 410 is gratefully acknowledged.
\end{acknowledgement}

\end{document}